\documentclass[aps,amssymb,amsmath, twocolumn, pra, superscriptaddress]{revtex4}
\usepackage{graphicx}
\usepackage{epsfig}
\usepackage{amsmath}
\usepackage{bm}
\usepackage{braket}
\usepackage{natbib}
\usepackage{enumitem}
\usepackage{longtable}
\usepackage{soul}

\usepackage[colorlinks, citecolor= magenta]{hyperref}
\usepackage[up]{subfigure}
\DeclareMathOperator{\Tr}{Tr}
\usepackage{epstopdf}

\usepackage{booktabs}

\usepackage{amsthm}

\newcounter{rtaskno}

\begin{document}

\title{Non-Markovianity of qubit evolution under the action of spin environment }
\author{Sagnik Chakraborty}
\email{csagnik@imsc.res.in}
\affiliation{Optics and Quantum Information Group, The Institute of Mathematical
Sciences, C. I. T. Campus, Taramani, Chennai 600113, India}
\affiliation{Homi Bhabha National Institute, Training School Complex, Anushakti Nagar, Mumbai 400094, India}
\author{Arindam Mallick}
\email{marindam@imsc.res.in}
\affiliation{Optics and Quantum Information Group, The Institute of Mathematical
Sciences, C. I. T. Campus, Taramani, Chennai 600113, India}
\affiliation{Homi Bhabha National Institute, Training School Complex, Anushakti Nagar, Mumbai 400094, India}
\author{Dipanjan Mandal}
\email{mdipanjan@imsc.res.in}
\affiliation{ The Institute of Mathematical
Sciences, C. I. T. Campus, Taramani, Chennai 600113, India}
\affiliation{Homi Bhabha National Institute, Training School Complex, Anushakti Nagar, Mumbai 400094, India}
\author{Sandeep K. Goyal}
\email{skgoyal@iisermohali.ac.in}
\affiliation{Department of Physics, Indian Institute of Science Education and Research, Mohali, Punjab, 140306 India}
\author{Sibasish Ghosh}
\email{sibasish@imsc.res.in}
\affiliation{Optics and Quantum Information Group, The Institute of Mathematical
Sciences, C. I. T. Campus, Taramani, Chennai 600113, India}
\affiliation{Homi Bhabha National Institute, Training School Complex, Anushakti Nagar, Mumbai 400094, India}

\begin{abstract}
The question, whether an open system dynamics is Markovian or non-Markovian can be answered by studying the direction of the information flow in the dynamics. In Markovian dynamics, information must always flow from the system to the environment. If the environment is interacting with only one of the subsystems of a bipartite system, the dynamics of the entanglement in the bipartite system can be used to identify the direction of information flow. Here we study the dynamics of a two-level system interacting with an environment, which is also a heat bath, and consists of a large number of two-level quantum systems. Our model can be seen as a close approximation to the `spin bath' model at low temperatures. We analyze the Markovian nature of the dynamics, as we change the coupling between the system and the environment. We find the Kraus operators of the dynamics for certain classes of couplings. We show that any form of time-independent or time-polynomial coupling gives rise to non-Markovianity.  Also, we witness  non-Markovianity for certain parameter values of time-exponential coupling. Moreover, we study the transition from non-Markovian to Markovian dynamics as we change the value of coupling strength. 

\end{abstract}

\maketitle

\section{Introduction}

We rarely come across systems that are completely isolated from the surrounding world. Had it been the case, dealing with quantum mechanical systems would have been lot more easier. So, although arduous to deal with, real quantum systems are mostly open quantum systems $-$ a system interacting with an environment. In these situations, information exchange between system and environment becomes an essential feature. Information that has been previously transferred to the environment 
may come back and affect the system, and this may appear as a memory-effect on the system. When this information backflow from the 
environment is negligible we have a situation analogous to the discrete Markov process, where the instantaneous state of the system depends
solely on the immediately previous step, the system dynamics is called {\it memory-less} or {\it Markovian}~\cite{breuer,rivas}.
On the other hand, when this information backflow affects the system significantly i.e. when some long past history of
the system influences its present state, the system dynamics becomes retentive, and is called {\it non-Markovian}.

In recent years, non-Markovianity has been used as a resource in a number of information theoretic protocols, namely, channel discrimination~\cite{bylicka}, preserving coherence and correlation~\cite{man,franco1} and retrieving quantum correlations in both quantum and classical environments~\cite{bellomo,gonzalez,xu,franco2013dynamics,bellomo2011dynamics}. Non-Markovian effects also play important roles 
in areas ranging from fundamental physics of strong fields \cite{schmidt,bloch} to energy transfer
process of photosynthetic complexes~\cite{rebentrost}. 

Owing to its diverse applications, various aspects of non-Markovianity are now being studied. Lately, researchers have been focusing on transition from non-Markovian to Markovian dynamics \cite{liu,bernardes,brito,garrido,man1,franco2}.
Some of them have dealt with bosonic bath of infinite or finite degrees of freedom, while some have considered a qudit system as the environment. But in all of these studies, system-environment interaction has been considered to be homogeneous in space, and the issue of non-Markovian to Markovian transition in terms of system-environment coupling strength has not been addressed. Note that, non-Markovian to Markovian transition is, in general, not a trivial issue, as in most cases finite dimensional environments give rise to non-Markovianity. 

In our study, we attempt to analyze the problem of whether a transition from non-Markovianity to Markovianity can be engineered for the spin bath model \cite{spinbath}. We particularly choose the spin bath model since it has wide ranging applications in simulating real physical scenarios 
\cite{hutton2004mediated, breuer2004non,spinbath}. In our attempt, we face a serious difficulty in diagonalizing the spin bath Hamiltonian, either analytically or numerically, for larger number of spins in the environment. Although, analytic solutions do exist for constant coupling \cite{samyadev} and some special forms of  time dependent coupling \cite{jing}, general solution for arbitrary forms of system-environment coupling of the spin bath Hamiltonian are hard to find. We therefore, try to circumvent the problem by choosing a simple model, which we argue, is a close approximation to the spin bath model for low temperatures. We choose an exchange type of interaction between a system qubit and individual environment qubits, where for each environment qubit the coupling can be chosen to be of different time dependent forms. But unlike the spin bath case, in our model, when the exchange interaction takes place between the system and a particular environment qubit, the rest of the environment qubits remain in a ground state; which also closely resembles the state of environment for low temperatures. As we will see in the paper, this approximation helps us to calculate and analyze non-Markovian to Markovian transition for
different types of system-environment coupling.

We present four scenarios here, for different forms of system-environment coupling: (i) the coupling is time-independent and homogeneous over environment qubits, (ii) the coupling is time-independent but inhomogeneous over environment qubits, (iii) the coupling is homogeneous over the environment but is time-dependent, and (iv) the coupling is both time-dependent and inhomogeneous. We find that cases (i) and (ii) always give rise to non-Markovian system dynamics. For cases (iii) and (iv), we find that some functional forms of coupling for certain ranges of coupling strengths gives rise to non-Markovianity. For example in case (iii), polynomial forms of coupling always give rise to non-Markovian system dynamics, while exponential coupling give rise to non-Markovian system dynamics only for certain ranges of parameter values. In case (iv) we find that a cross-over from non-Markovianity to  Markovianity can be achieved by varying the strength of coupling. We also calculate, the extremal values  of coupling parameter beyond which non-Markovianity can no longer be detected. Thus we see, these extremal values act as critical values for transition from  non-Markovian to Markovian regime. It is worth mentioning here that, for the purpose of detecting non-Markovianity we use  Rivas-Huelga-Plenio (RHP) 
measure of non-Markovianity as proposed in \cite{rivas2}. Although there are different approaches of defining Markovianity  and each approach represent different aspects of Markovianity, for the purpose of the present paper we choose, detection by the 
 RHP measure as  the definition of Markovianity.

Similar works on this line were done in \cite{apol1,apol2,wang2013non}. But in the first approach \cite{apol1}, the system qubit directly interacts with a single environment qubit and the rest of the environment qubits, only have an indirect effect on the system via the environment qubit directly attached. Also, the coupling parameters involved do not have any time dependence. In the second approach \cite{apol2}, the transition from Markovianity to non-Markovianity was shown with a two tier environment; the first one being a multiple-spin system, while the second one was a bosonic bath. Also in \cite{wang2013non}, the coupling between the system and individual environment qubits were constant in space and time. We take into account all these factors and present a detailed study of a spin environment and cover {\it all} the relevant cases.

In section \ref{background}, we discuss the relevant background required for following the techniques used in the paper. In section \ref{Sec:Model}, we present our model and in section \ref{Sec: system-bath} we introduce different types of couplings and theredy analyze them. Finally, in section \ref{results}, we present the results of our analysis, before concluding in section \ref{conclu}.

\section{Background}\label{background}
In this section we present the relevant background of Markovian dynamics and the definitions used in the  paper. We also describe the measure of entanglement for two-qubit systems, which will also be used to quantify non-Markovianity of our dynamics.

\subsection{Quantum Markovian dynamics}\label{evolut}
A discrete time stochastic process is called Markovian (Markov chain) if the state of the system at time $t_n$ depends solely on the state of the system at time $t_{n-1}$. This concept of Markov chain can be extended to the continuous time stochastic processes as well~\cite{rivas}. However, generalizing  it to quantum dynamics is a difficult task. Numerous prescriptions have been proposed to capture different aspects of quantum Markovianity. Broadly these prescriptions can be classified into two classes: information backflow~\cite{breuer1,wise,luo,bylicka2017constructive,chruscinski2017universal, chakraborty} and completely positive divisibility (CP-divisibility)~\cite{breuer,chru}.

{\it Information backflow:} The information backflow approach is inspired from the fact that a Markovian dynamics is characterized by unidirectional flow of information from the system to the environment. As for example, in the Lindblad master equation \cite{breuer2002theory}, the non-negativity of the entropy production rate signifies unidirectional information flow from system to the environment, and thereby, is a signature of Markovianity. A dynamics is called  Markovian from the information backflow approach, if some information quantifier decays over time in a monotonic way. Any departure from monotonicity of such quantifier is seen as a backflow of information from the environment, back to the system. Different quantifiers of information like distinguishability of states \cite{breuer1}, measure of entanglement \cite{rivas2}, quantum mutual information \cite{luo}, etc has also been suggested for this purpose. Each quantifier provides a different definition of Markovianity; all of which, are not in general equivalent. Only recently, there has been attempts to unify all these different definitions \cite{bylicka2017constructive,chruscinski2017universal,chakraborty} to provide a unified approach to information backflow. 

{\it CP-divisibility:} Any dynamical process, given by  completely-positive (CP) trace preserving (TP) map $\Lambda_t$ representing evolution up to time $t$ is called CP-divisible if
\begin{equation}
\label{cpd}
 \Lambda_t=V_{t,s}\circ\Lambda_s,
\end{equation}
where $V_{t,s}$ is CP for any $t \ge s\ge 0$, and $\circ$ denotes composition.

Although the most general description of Markovian dynamics is given by CP-Divisibility \cite{breuer,chru}, for our purpose we consider information backflow, in terms of measure of entanglement i e. the RHP measure, as the description of Markovianity.

\subsection{Detecting non-Markovianity through Entanglement}
Let us first discuss entanglement measure of two qubit states. The entanglement between two two-level systems (two qubits) can be characterized by the Peres-Horodecki criterion~\cite{peres,horod} which states that a two-qubit state $\rho_{\text{as}}$,  shared between a system qubit $s$ and an ancilla qubit $a$, is entangled if and only if the partial transpose of this state, i.e. $( \rho_{\text{as}} )^{T_s}$, is  not a positive-semidefinite operator i.e. $( \rho_{as} )^{T_s}\ngeq0$. Notably, for a two-qubit entangled state, the operator $( \rho_{\text{as}} )^{T_s}$ has exactly one negative eigenvalue $\lambda$ \cite{sanpera1998local, rana2012entanglement}. Thus $|\lambda|$ may be used as a measure of entanglement for the state $\rho_{\text{as}}$.
Formally, the entanglement measure can be defined as follows
\begin{equation}\label{ent-measure}
 E(\rho_{as})=\frac{||(\rho_{as} )^{T_s}||_1-1}{2}
\end{equation}
where, $||A||_1=\Tr \sqrt{A^{\dagger}A}$ is the trace norm of a matrix $A$. Note that, $E(\rho_{as})$ is nothing but the negativity of the bipartite state $\rho_{as}$ \cite{vidal2002computable}. We will use this measure of entanglement as the quantifier for ascertaining Markovianity of the dynamics from the information backflow approach. Using entanglement to detect non-Markovianity was first used by Rivas, Huelga and Plenio in \cite{rivas2}, and this measure has been so called the RHP measure of non-Markovianity. Following their technique we attach an ancilla to the system, on which a dynamical map $\Lambda_t$ is acting. Following the information backflow approach, the dynamical map $\Lambda_t$ is called Markovian if $E\big((\openone\otimes\Lambda_t)\big[\ket{\Phi^+}\bra{\Phi^+}\big]\big)$ is a non-increasing function of time $t$, where $\ket{\Phi^+}\bra{\Phi^+}$ is the maximally entangled state, given by
\begin{equation}
\label{max_entangled_state}
  \ket{\Phi^+} = \frac{1}{\sqrt{2}} \big(\ket{00} + \ket{11}\big).
 \end{equation}



\section{The model} \label{Sec:Model}
\begin{figure}
 \includegraphics[width=\columnwidth, height=4.5 cm]{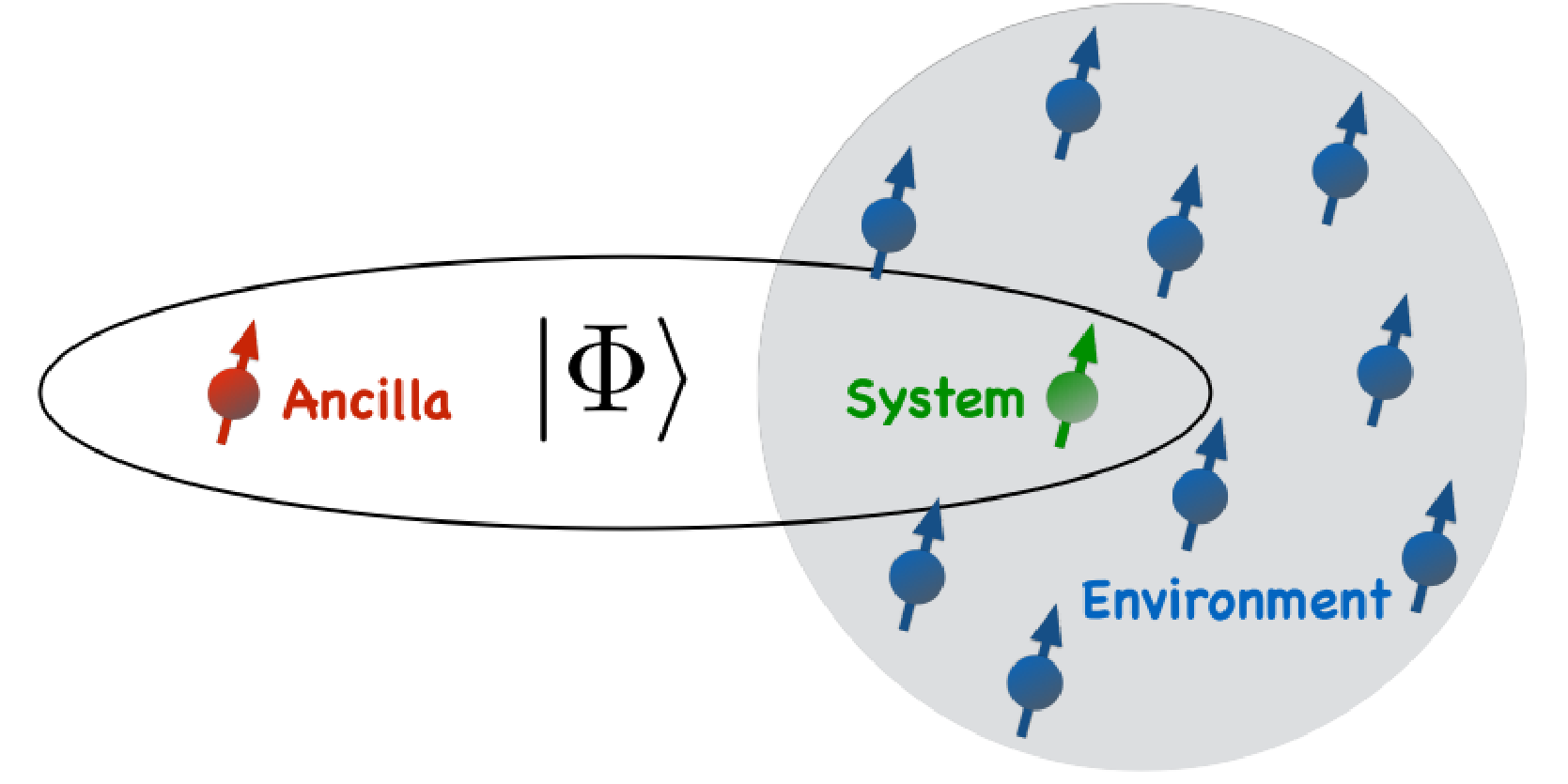}
 \caption{(colour online) Schematic diagram of system qubit and ancilla qubit sharing a maximally entangled state $\ket{\Phi}=\frac{1}{\sqrt{2}}(\ket{11}+\ket{00})$. The system is 
 interacting with an environment consisting of finite number of non-interacting qubits.}
 \label{fig10}
 \end{figure}

In this section, we present our model and discuss the motivation behind choosing it. We also describe the technique in detail, in which non-Markovianity in the system dynamics is detected. We consider two qubits, one of which is called the system ($s$) and the other, the ancilla ($a$). The system qubit is placed in an environment consisting of $N$  non-interacting qubits (see Fig.~\ref{fig10}). 
We take the interaction between the system qubit and the environment in the following form,
\begin{eqnarray}\label{hamil_original}
\tilde H_{se}(t)&=&\hbar \alpha  \Big[ \ket{1}_s\bra{0} \otimes\sum_{n=1}^N \tilde g^*_n(t) \ket{0..0_n..0}_e\bra{ 0..1_n..0}  \nonumber\\
 &+&   \ket{0}_s\bra{1} \otimes\sum_{n=1}^N \tilde g_n(t) \ket{0..1_n..0}_e\bra{ 0..0_n..0}  \Big],
\end{eqnarray}
where $\ket{0}$ and $\ket{1}$, respectively represent the ground and excited states of each qubit.
The coupling strength $\tilde g_n(t)$ is in general, a complex number and can also be time-dependent as well as site-dependent, and $ \alpha$ is a real parameter with the dimension of frequency. The extra factor $\alpha$ is introduced to make the coupling strengths $\tilde g_n(t)$ dimensionless. For all practical purposes $\alpha$ can be assumed to be $1$. 
The free Hamiltonians of the system and the environment are respectively given by,
\begin{align}
  H_s & = \frac{\hbar\omega_s}{2} \sigma_z,\\
  H_e & = \sum_n \frac{\hbar\omega_n}{2} \sigma^{(n)}_z.
\end{align}
It is convenient to work in the interaction picture where we replace the total Hamiltonian $H = H_s + H_e + \tilde H_{\text{se}}\equiv H_0 + \tilde H_{\text{se}}$ by the interaction picture  Hamiltonian $H_{\text{se}}(t) = \exp(i H_0 t/\hbar)\tilde H_{\text{se}}\exp(-iH_0 t/\hbar)$ which reads,
\begin{align}\label{hamil}
  H_{\text{se}}(t) &=\hbar \alpha  \Big[ \ket{1}_s\bra{0} \otimes\sum_{n=1}^N \tilde g^*_n(t) e^{i\delta\omega_n\, t} \ket{0..0_n..0}_e\bra{ 0..1_n..0}  \nonumber\\
  &+   \ket{0}_s\bra{1} \otimes\sum_{n=1}^N \tilde g_n(t)e^{-i\delta\omega_n\, t} \ket{0..1_n..0}_e\bra{ 0..0_n..0}  \Big]\nonumber\\
  &=\hbar \alpha  \Big[ \ket{1}_s\bra{0} \otimes\sum_{n=1}^N g^*_n(t) \ket{0..0_n..0}_e\bra{ 0..1_n..0}  \nonumber\\
  &+   \ket{0}_s\bra{1} \otimes\sum_{n=1}^N g_n(t) \ket{0..1_n..0}_e\bra{ 0..0_n..0}  \Big]
\end{align}
where $\delta\omega_n = \omega_s-\omega_n$ and $g_n(t) = \tilde g_n(t)e^{-i\delta\omega_n}$. Henceforth, our discussion will be based on the Hamiltonian $H_{se}(t)$. We also consider the initial state of the environment to be in the thermal state,
 \begin{equation}\label{rhoe}
 \rho_e(0) = \Big[p \ket{0}\bra{0}+(1-p)\ket{1}\bra{1}\Big]^{\otimes N},
 \end{equation}
 where $p=\big(1 + e^{- \beta}\big)^{-1}$ and $ \beta $ is a positive real parameter which can be identified as the inverse of the temperature $T$ of the environment. 

\subsection{Motivation behind the model}
 Here we argue that, our model is   a close approximation to the `spin bath' model \cite{hutton2004mediated, breuer2004non} at low temperatures. Note that the Hamiltonian in Eq. (\ref{hamil}) can also be written as, 
%
\begin{align}\label{hamil1}
  H_{\text{se}}(t) &=\hbar \alpha  \sum_{n=1}^N \Big\{g^*_n(t)\sigma_+^{(s)}\otimes\big[\ket{0}\bra{0}\otimes.. \sigma_-^{(n)}..\otimes\ket{0}\bra{0}\big]_e\nonumber\\
  &+g_n(t)\sigma_-^{(s)}\otimes\big[\ket{0}\bra{0}\otimes.. \sigma_+^{(n)}..\otimes\ket{0}\bra{0}\big]_e\Big\}
\end{align}
where $\sigma_+=\ket{0}\bra{1}$ and $\sigma_-=\ket{1}\bra{0}$. When we compare Eq. (\ref{hamil1}) with the usual Hamiltonian of a spin bath model \cite{hutton2004mediated, breuer2004non, samyadev} in the interaction picture, given by,
\begin{equation}
 H_{spin-bath}=\hbar\alpha\sum_{n=1}^N\big(\sigma_x^{(s)}\sigma_x^{(n)}+\sigma_y^{(s)}\sigma_y^{(n)}+\sigma_z^{(s)}\sigma_z^{(n)}\big)
\end{equation}
we find that the only difference comes from the $\ket{0}\bra{0}$ factors arising in Eq. (\ref{hamil1}), which are replaced by $\openone$ for the spin bath Hamiltonian. As a result of this difference, the dynamics of the spin bath model is not entirely the same as our model. In the former, an exchange of one quanta of energy takes place between the system and individual environment qubit, when the rest of the environment qubits are allowed to be in any state, whereas in the later, the exchange will only take place when the rest of the environment qubits are in their ground state. This difference, although significant in general, will not play a major role when the state of the environment is close to the ground states, or in other words,  temperature of the environment is low. Note that low temperature of environment correspond to values of $p$ in Eq. (\ref{rhoe}), which are very close to $1$, and this also confirms the fact that for low temperatures $\rho_e$ is close to the ground state. Thus we see for low temperatures our model serves as a close approximation to the spin bath model. The main advantage of our model 
is the fact that our Hamiltonian is easily diagonalizable, and for certain types of couplings, as we discuss later in detail, allows for exact determination of the system dynamics in terms of Kraus operators, for {\it any} number of environment qubits.

We also stress that, although our model shows similarity to the spin bath model for low temperatures, we find solutions and analyze the dynamics of our model for any temperature whatsoever. The reason behind this is that our model being analytically solvable for certain types of couplings, allows for an opportunity to exactly solve the dynamics for any number of environment qubits, which is not often the case for systems with large number of spins. Note that, even for the spin-bath Hamiltonian, it is not easy to find the exact solution for non-zero temperature.

\subsection{Diagonalizing the Hamiltonian of our model}

There are only two non-zero eigenvalues of the Hamiltonian $H_{\text se}(t)$ and they are,
\begin{align}\label{nonzeroeigen} \mathcal{E}_\pm(t) 
= \pm \hbar \alpha   \sqrt{ \sum_{n=1}^N |g_n(t)|^2}=\pm\mathcal{E}(t),
\end{align}
corresponding to the eigenvectors,
\begin{align}\label{non-zero}
 \ket{\chi_\pm(t)}_{se} = \frac{1}{\sqrt{2}} \big[ \ket{1}_s \otimes \ket{0}^{\otimes N}_e \pm  \ket{\xi(t)}_{se}  \big],  
\end{align}
where $\ket{\xi(t)}_{se} = \ket{0}_s\otimes\ket{\beta_0}_e$ and,
\begin{equation}
 \label{beta0}
 \ket{\beta_0}_e=\frac{\hbar \alpha}{\mathcal{E}(t) } 
\sum_{n=1}^N g_n(t) \ket{ 0..1_n..0}_e.
\end{equation}
Thus, the time evolution operator $U(t,0)$ corresponding to the Hamiltonian $H_{\text se}$ is,
\begin{align}
  U(t,0) = \mathcal{T}\exp\left[ - \frac{i}{\hbar} \int_{0}^t H_{se}(\tau') ~ d\tau' \right],
\end{align}
where $\mathcal{T}$ represents time ordering. 

The ancilla qubit is used as a probe to characterize the non-Markovianity of the dynamics of the system in the presence of the environment. In order to do so we prepare the system and ancilla qubits in a maximally entangled state $\Ket{\Phi^+}$, as given in Eq. (\ref{max_entangled_state}).
Due to the interaction of the system qubit with the environment, the entanglement between the system and the ancilla qubit will evolve with time. The deviation of this time evolution of the entanglement, from monotonic decay is used to establish the non-Markovian character of the dynamics. Note here, that this idea was used by Rivas et al \cite{rivas2} to devise a measure of non-Markovianity. In the present paper, we follow this technique to consider the system dynamics to be non-Markovian whenever the entanglement between system and ancilla, as described above, shows non-monotonic behaviour, otherwise we consider the dynamics to be Markovian.

The joint initial state of the system plus ancilla plus environment is of the form, 
\begin{equation}
  \rho_{\text {ase}}(0) = \Ket{\Phi^+}\Bra{\Phi^+} \otimes \rho_e(0),
  \label{rhoase}
\end{equation}
which evolves to,
\begin{equation}\label{wholefinal}
 \rho_{ase}(t) = \big[ \mathbb{I}_a \otimes U(t,0) \big] \rho_{\text{ase}}(0) \big[ \mathbb{I}_a \otimes U^{\dagger}(t,0) \big].
\end{equation}
 Therefore, reduced time-evolved system-ancilla state can be calculated by tracing out the environment part, 
 \begin{equation}\label{finalsysan}
   \rho_{as}(t) =  \text{Tr}_e \,\rho_{\text{ase}}(t) .
 \end{equation}

 \section{System - Environment couplings}\label{Sec: system-bath}
In this section, we introduce various classes of system-environment coupling, and in each case, we study their effect on the evolution of the system-ancilla joint state. We classify all the couplings into four major classes : (A) when the coupling parameter $g_n(t)$ is independent of the site index $n$ (homogeneous) and time-independent; (B) when $g_n(t)$ is inhomogeneous but time-independent; (C) when $g_n(t)$ is homogeneous but time-dependent, and (D) when $g_n(t)$ is inhomogeneous and time-dependent. For each class, we calculate the entanglement of the time evolved state of system-ancilla, and thereby try to characterize the non-Markovian behaviour of the system dynamics. Henceforth, we assume $\alpha$ to be $1$.

\subsection{Homogeneous and time-independent coupling}\label{subsec0}
We have here the simplest situation, where the coupling of the system with all the environment qubits are uniform and time-independent i.e. $g_n(t)=g$, a constant. As a result, the non-zero eigenvalues of the Hamiltonian, as given in Eq. (\ref{nonzeroeigen}), takes the form  $\mathcal{E}_{\pm}(t)=\pm\mathcal{E} = \pm \hbar \sqrt{N}~|g| \equiv \hbar \omega_0$, where $\omega_0= \sqrt{N}~|g|$ is a constant with the dimension of frequency. The time-evolution operator $U(t,0)$ is of the form,
\begin{align}\label{unit} 
U(t,0)=& \left(e^{-i\omega_0 t}-1\right)\ket{\chi_+}\bra{\chi_+} \nonumber\\
   &\quad+ \left(e^{i\omega_0 t}-1\right)\ket{\chi_-}\bra{\chi_-} + \openone.
\end{align}
Using the above form and the form of $\rho_e$ given in Eq. (\ref{rhoe}), we find the Kraus operators $K_{mn}(t)$ of system dynamics, which are defined in the following way,
\begin{equation}
 \rho_s(0)\rightarrow\rho_s(t)=\sum_{m,n=1}^N K_{mn}(t)~\rho_s(0)~K_{mn}^{\dagger}(t),
\end{equation}
where the $N^2$ Kraus operators are given by,
\begin{widetext}
\begin{equation}
 K_{mn}(t)=\sqrt{p^{N-s_n}(1-p)^{s_n}}\left[
 \begin{array}{c c c}
  \frac{\cos \omega_0 t-1}{\omega_0^2} ~g_{N-\log m}(t)~g^*_{N-\log n}(t)+\delta_{mn} && -\frac{i}{\omega_0}\sin \omega_0 t ~g_{N-\log m}(t)~\delta_{0n}\\ \\
  -\frac{i}{\omega_0}\sin \omega_0 t~ g^*_{N-\log n}(t)~\delta_{0m} && (\cos\omega_0t-1)~\delta_{0m}\delta_{0n}+\delta_{mn}\\
 \end{array}\right]
 \label{krausform}
\end{equation} 
\end{widetext}
where $m,n=1,\dots,N$, $\log x$ refers to $\log_2 x$, and $s_n$ is the number of $1$'s in the binary equivalent of $n$. For example, if $n=6$, then the binary equivalent of $n$ is $110$. Therefore $s_n=2$. 

We then find time evolved state of the system-ancilla, using Eqs. (\ref{rhoase}), (\ref{wholefinal}) and (\ref{finalsysan}),
\begin{align}\label{eigneg0}
  \rho_{\text{as}}(t) =& \Ket{\Phi^+}\Bra{\Phi^+}-\frac{1}{2}\Big[ p\kappa_0 \big(\ket{11}\bra{11} +  \ket{10}\bra{10}\big) \nonumber\\
  &+ (1-p) \kappa_0 \big(\ket{00}\bra{00} +\ket{01}\bra{01}\big)  \nonumber\\
 &+  \delta_0( \ket{00}\bra{11}+ \ket{11}\bra{00})\Big],
\end{align}
where $\kappa_0 = p^{N-1} \sin^2 (\omega_0 t)$ and $\delta_0 =2 p^{N-1} \sin^2 \left( \frac{\omega_0 t}{2}\right)$. The only possible negative eigenvalue of $  [ \rho_{\text{as}}(t) ]^{T_s} $, if any, is of the form,
\begin{align}
\label{consenta0}
  \lambda(t) =\frac{1}{2}\left( \kappa_0 - \sqrt{(1-2p\kappa_0)^2 + 4\delta_0^2}\right).
\end{align}
We present the plot of $E(\rho_as(t))=\lambda(t)$ versus time, later in the Result section. 

\begin{figure*}     
     \includegraphics[width= 2.0\columnwidth]{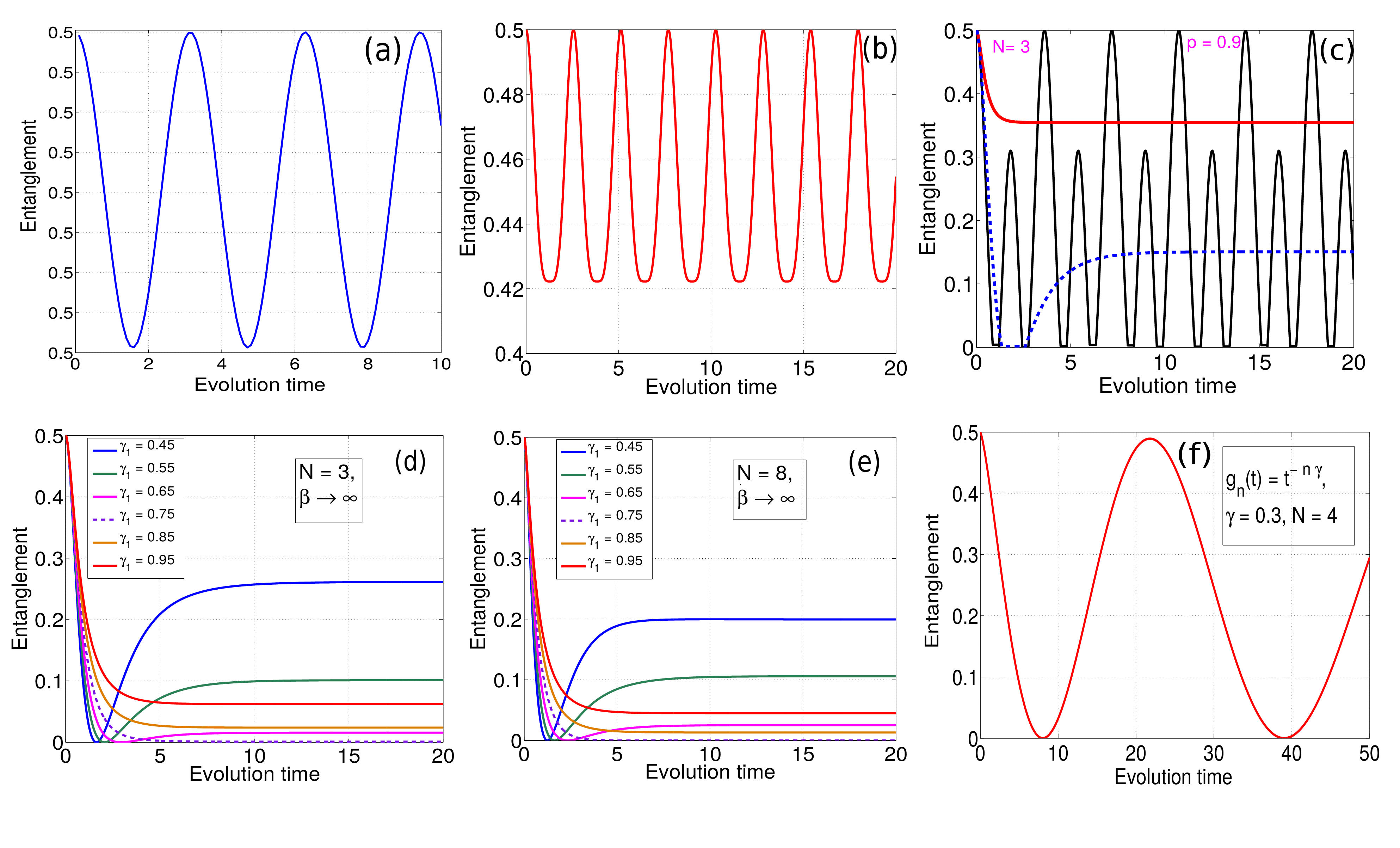}
     \caption{\small{(colour online) Plots showing the system-ancilla entanglement dynamics in different scenarios. For simplicity we have considered $\alpha = 1$. (a) When the coupling is homogeneous and time-independent for $N=6$ and $p = 0.6$. (b) When the coupling is inhomogeneous and time-independent, for $N=6$ and $p = 0.6$. (c) When the coupling is homogeneous and  time-dependent, i.e. $g_n(t) = g(t)$. Here we consider two different time-dependent functions $g(t)$. One is exponential in time: $g(t) = \exp(-\gamma t)$ and other in the form: $g(t) = t^{-\delta}$. The blue and the red curves represent the entanglement dynamics when  $g(t) = e^{ - 0.7 t}$ and $ g(t) = e^{- 2.7 t} $, respectively. The black curve is for the entanglement dynamics when $g(t) = t^{0.01}$.
  (d), (e) The coupling $g_n(t) = e^{ - \gamma_1 n t}$ is inhomogeneous and time-dependent. Here the number of qubits in the bath are $N=3$, and  $N=8$, respectively, and the environment is taken to be at zero temperature i.e. $p=1$.  We see that there is a transition   from non-Markovianity to Markovianity as the coupling strength is made weaker and weaker. (f) Another example of inhomogeneous time-dependent coupling. Here the initial temperature of the environment is set to zero, i.e. $p=1$.}}
   \label{comb}
\end{figure*}

\subsection{Inhomogeneous  and time-independent coupling}\label{subsec1}
Consider a system, where a single two-level system (perhaps an ion as an impurity) is placed in a spin lattice. The lattice sites, closest to the impurity interacts very strongly with the system, while, as we go away from the impurity site, the strength of interaction becomes weaker and weaker. In such cases the interaction parameter $g_n(t)$ is in general inhomogeneous, but there is no explicit time dependence. Therefore, $g_n(t) = g_n$. Hence, $\mathcal{E}_{\pm}(t)=\pm\mathcal{E} = \hbar \omega$ in Eq. (\ref{nonzeroeigen}) are also time-independent. Note, in this case also $\omega=\sqrt{ \sum_{n=1}^N |g_n|^2 }$ is a constant with dimensions of frequency. Following this, the analysis is same as in the last subsection. As a result, the evolution operator $U(t,0)$, the time evolved state $\rho_{\text{as}}(t)$ and the only possible negative eigenvalue, if any, of $  [ \rho_{\text{as}}(t) ]^{T_s} $ for this case are of the same forms as in Eqs (\ref{unit}), (\ref{krausform}), (\ref{eigneg0}) and (\ref{consenta0}) respectively, except for $\omega_0$, in appropriate places, replaced by $\omega$.


\subsection{Homogeneous and time-dependent coupling}\label{subsec2}
So far we have considered only couplings which are independent of time. In this section, we consider time-dependent and homogeneous couplings. We take an {\it arbitrary} real function of time, which is independent of site index $n$ i.e. $g_n(t) =g(t)$.
Note that our coupling operator between system and individual environment qubit, as given in Eq. (\ref{hamil}), is of the form $\sigma_+\otimes \sigma_- + \sigma_-\otimes \sigma_+$ which can also be expressed as $\sigma_x\otimes\sigma_x + \sigma_y\otimes
\sigma_y$. Thus, our system-environment coupling is a special case of the $XY$ coupled Hamiltonian. Such coupling with time-dependent coefficients have been used to show non-trivial entanglement dynamics~\cite{sadiek2010entanglement,bortz2007spin}.

Fortunately, the Hamiltonians $H_{se}(t)$ in this case commutes at different times, which makes the analysis similar to the one in section \ref{subsec1}. The only difference being the non-zero eigenvalues, in Eq.~(\ref{nonzeroeigen}), to be of the form $\mathcal{E}_\pm(t)= \pm \hbar   \sqrt{N}|g(t)|=\pm\mathcal{E}(t)$, which is no longer constant in time. The whole treatment of the dynamics of the system and the ancilla remains the same if we replace $\omega_0$ and $\omega_0 t$, in Eqs (\ref{unit}), (\ref{krausform}), (\ref{eigneg0}) and (\ref{consenta0}), by $\sqrt{N}~|g(t)|$ and $\Omega(t)$, respectively, where,
\begin{align}
 \Omega(t)&=\frac{1}{\hbar} \int_0^t  \mathcal{E}(\tau) d\tau =  \int_0^t   \bigg( \sum_{n=1}^N |g_n(\tau)|^2  \bigg)^{\frac{1}{2}} d\tau \nonumber\\
 &= \sqrt{N} \int_0^t |g(\tau)| d\tau.
\end{align}

\subsection{Inhomogeneous and time-dependent coupling}\label{subsec3}

The most general class of coupling  $g_n(t)$ is when it depends on both the site $n$ and time $t$. The interaction  Hamiltonian in such a situation  does not commute at different times and this makes the calculation for solving the dynamics difficult. However, we can use numerical methods to simulate the time-evolution and get the solution for $\rho_{\text{as}}(t)$.


One can obtain the following results analytically before starting the simulation part.

{\em Analytical Part:}  Two of the eigenvalues of the Hamiltonian in Eq. (\ref{hamil}) are non-zero, as given in Eq. (\ref{nonzeroeigen}). The remaining $( 2^{N+1}-2 )$ of the eigenvalues are zero. A possible choice for these null space eigenvectors are found in the following way:

{\bf Step I:} We first feed the eigenvectors (corresponding to non-zero eigenvalues) given in Eq. (\ref{non-zero}) as rows of a matrix $A$. Note, $A$ is a $2\times 2^{(N+1)}$ matrix

{\bf Step II:} By row reduction method \cite{hoffmanlinear}, we find out a basis $\mathcal{B}$ for the Null space of $A$. Note that $\mathcal{B}$ is not necessarily ortho-normal.

{\em Simulation Part:} Obtaining an orthonormal basis $\mathcal{B}'$ from $\mathcal{B}$ analytically is a challenging job. We, therefore resort to numerical techniques for this case.

{\bf Step I:} From $\mathcal{B}$, using Gram-Schmidt Orthonormalization procedure \cite{hoffmanlinear}, we find an orthonormal basis $\mathcal{B}'$. Note, $\mathcal{B}'$ forms the set of eigenvectors of the  Hamiltonian, corresponding to zero eigenvalues.  

{\bf Step II:} As, the eigenvectors are time-dependent, the Hamiltonian is not different-time commuting. Hence, the evolution operator may be found numerically from the  following expression,
  \begin{eqnarray}\label{unitaryform} 
U_{se}(t,0)&=&\Large{\mathcal{T}}exp\Big[ - \frac{i}{\hbar} \int_{0}^t H_{se}(\tau') ~ d\tau' \Big]\nonumber\\
&=&\lim_{m \to \infty}\Bigg[ exp\Big[ - \frac{i}{\hbar} \int_{(m-1)\tau}^{m\tau} H_{se}(\tau') ~ d\tau' \Big]\nonumber\\
&\times& exp\Big[ - \frac{i}{\hbar} \int_{(m-2)\tau}^{(m-1)\tau} H_{se}(\tau') ~ d\tau' \Big]\times...\nonumber\\
&\times& exp\Big[ - \frac{i}{\hbar} \int_{0}^{\tau} H_{se}(\tau') ~ d\tau' \Big]\Bigg],
 \end{eqnarray}
where $\mathcal{T}$ represents time ordering.
 
{\bf Step III:} We evolve the initial ancilla -system-environment state $\rho_{ase}(0)$ by the unitary operator $U_{ase}(t,0)=\mathbb{I}_a\otimes U_{se}(t,0)$ and get the time evolved state  $\rho_{ase}(t)$.

{\bf Step IV:} We trace out the environment from $\rho_{ase}(t)$ and get $\rho_{as}(t)=Tr_e[\rho_{ase}(t)]$. We then evaluate our entanglement measure $E(t)$ given in Eq. (\ref{ent-measure}), on $\rho_{as}(t)$ and plot it as a function of time.

\begin{figure}[]
  \includegraphics[width= 1.0\columnwidth]{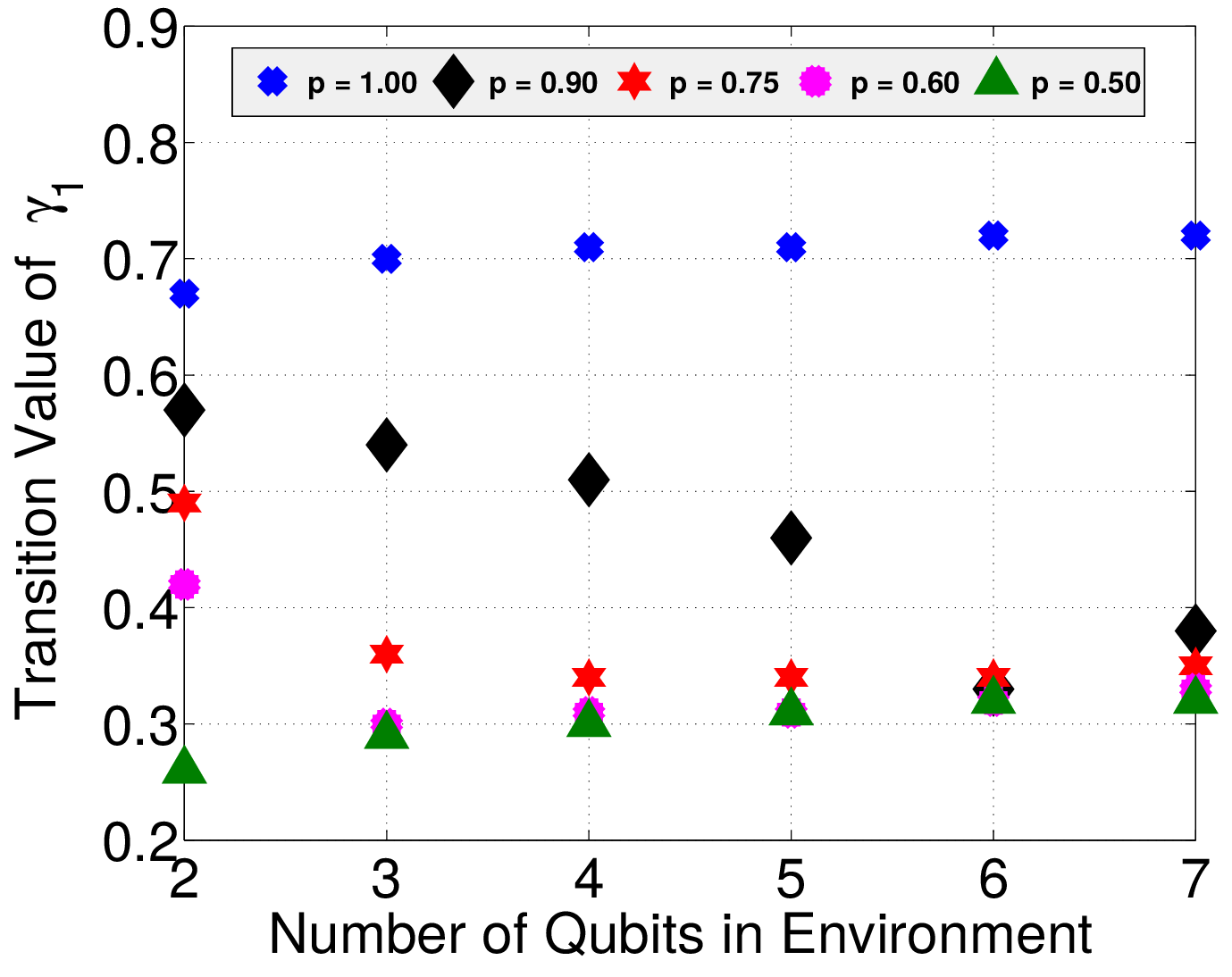}
\caption{\small{(colour online) Plot showing the transition values of $\gamma_1$ for coupling $ g_n(t) = e^{ - \gamma_1 n t}$, as a function of number of environment qubits $N$ for different values of environment temperature $p$.
}}
\label{trans}
\end{figure}

\section{Results}\label{results}

In this section, we show that some of the classes of the couplings that we have considered, always results in non-Markovian dynamics. However, there are also some classes for which we can tune the parameters and find a transition from non-Markovian to Markovian dynamics. In order to do so, we plot the entanglement dynamics between the system and ancilla, for each class,  as a function of time and observed if there is any departure from monotonicity with time. As mentioned in Sec \ref{Sec:Model}, this technique helps in characterizing non-Markovianity present in the system dynamics. In Fig.~\ref{comb}, we present entanglement dynamics for different classes of the system-environment coupling, considered in the previous section. We now present our findings for each class of system-environment coupling.

{\it Section \ref{subsec0} and \ref{subsec1}} : In Figs.~\ref{comb}(a) and \ref{comb}(b), we plot the entanglement as a function of time for the homogeneous time-independent, and the inhomogeneous time-independent couplings, respectively, i.e, $g_n(t) = g$ and $g_n(t) = g_n$, respectively. Note here, that $g$ and $g_n$, for all values of $n$, are arbitrary complex functions. Analytic calculations for the entanglement measure, as given in Eq. (\eqref{consenta0}) suggests a periodic behaviour for both the classes, which can also be seen in Figs.~\ref{comb}(a) and \ref{comb}(b). As a result, we conclude in both of these classes of couplings, the dynamics is always non-Markovian.

{\it Section \ref{subsec2}} : We consider homogeneous and time-dependent couplings and find that if $g(t)$ is some polynomial function of $t$, we will get $\Omega(t)$ as a polynomial function of t. This gives rise to a periodic function $\lambda(t)$. As a result, the dynamics is non-Markovian, in general. Interestingly, it was recently pointed out \cite{dive2015quantum}, that the Hamiltonian dialation obtained from the Choi state of the system dynamics, diverges whenever the system dynamics is time-independent Markovian. Although, in our case, the system-environment Hamiltonian (equivalent to the dialated Hamiltonian)is not related to the Choi state of the system dynamics in the same way as in \cite{dive2015quantum}. As a result the conclusion of \cite{dive2015quantum} differs from ours.
If $g(t) = \exp(-\gamma t)$ then non-Markovianity can be witnessed if the real part $\gamma_r$ of $\gamma$ fails to be positive or violates the inequality $\alpha \sqrt{N}\ge\gamma_r \pi$. Fig.~\ref{comb}(c) shows the entanglement vs time plot for two values of $\gamma_r$; one of which violates the above mentioned inequality. We also consider the case $g(t) = t^{-\delta}$, and show that the dynamics is non-Markovian for $\delta=0.01$.

{\it Section \ref{subsec3}} : For inhomogeneous time-dependent couplings, the dynamics can be made both non-Markovian and Markovian by choosing the strength of the coupling appropriately. We consider two special cases of inhomogeneous time-dependent coupling: $(i)$ $ g_n(t) = e^{ - \gamma_1 n t}$, and $(ii)$ $ g_n(t) = t^{-n \gamma} $. For simplicity, we have assumed $ \alpha = 1$ and set the temperature of the environment to be zero i.e. $\beta \to \infty$. For analyzing coupling $(i)$, we plot the system-ancilla entanglement measure as a function of time in Figures \ref{comb}(d), and \ref{comb}(e), for different values of the coupling parameter $ \gamma_1$, at a fixed value of $N$ i.e. number of qubits in the environment. In Figures \ref{comb}(d), and \ref{comb}(e), monotonically decreasing entanglement values show signs of  Markovianity and non-monotonic decay are evidence of non-Markovianity. As expected, increasing the coupling parameter  $ \gamma_1 $ i.e., decreasing coupling strength, leads to the transition from non-Markovian to Markovian dynamics. The figures also show an interesting feature that, after sufficient time, the entanglement in the system-ancilla state saturates to fixed values irrespective of their Markovian or non-Markovian nature. This feature can be signs of possible equilibration of the system ancilla state. For coupling $(ii)$, the dynamics showed non-Markovianity for $\gamma=0.3$, as shown in Fig \ref{comb}(f).

Next, for coupling $(i)$ i.e. $ g_n(t) = e^{ - \gamma_1 n t}$, we find the extremal values of $\gamma_1$ for which non-Markovianity is witnessed. These extremal values serve as transition parameters from non-Markovianity to Markovianity. On plotting these transition values as a function of $N$ (Fig.~\ref{trans}) it appears that a saturation is reached as $N$ is increased. 

 \section{ Conclusion}\label{conclu}
 In this paper, we have addressed the question of how non-Markovianity of a dynamics changes with the interaction between the system and the environment and also with size of the environment. We have taken a simple model constituting of a few qubits, which can also be seen as a  close approximation to the spin bath model, for low temperatures. Even in this minimalistic scenario, we were able to find a transition from non-Markovian to  Markovian dynamics by tuning the system-environment interaction. This is somewhat counterintuitive as it is generally conceived that for having Markovian dynamics the bath/environment should have infinite degrees of freedom, although there are exceptions \cite{pang2017abrupt}. We also found, in our model, that if the interaction Hamiltonian is time-independent, the dynamics is always non-Markovian, irrespective of the size of environment. Note that in this scenario for a general interaction Hamiltonian in the weak coupling limit, we generally get to see Markovian dynamics only for a very large size of bath; at least in the case of harmonic oscillator bath. The present scenario is  different as we have considered spin environment and the system-environment interaction is of very specific type. In the case of site-independent interaction, polynomial forms and certain cases of exponential forms of interaction show non-Markovianity. Lastly, we study time-dependent and site-dependent interaction for certain forms of system-environment coupling. In this last case, we also saw a transition from non-Markovian to  Markovian regime. Interestingly, the transition values appear to saturate to a certain value depending on the initial temperature of the environment, as the number of environment qubits increases.
 Examining this type of spin environment is recently drawing some amount of interest~\cite{spinbath}. Studies on similar lines was also done recently in \cite{samyadev}, where an analysis of a qubit system interacting with a sea of spins was given. A number of
 questions arise from the present study: whether such a transition can be found by considering more general forms of interaction, what happens if 
 along with system-environment interaction there is some interaction present among the environment particles themselves, etc. One may also find it useful
 to check, whether the aforesaid saturation of transition parameters is a general feature of interaction that exhibits non-Markovian to Markovian transition.
 
\begin{acknowledgments}
The authors would like to thank R. Lo Franco for helping in revising the earlier version of the manuscript as well as providing the references \cite{franco1,bellomo,gonzalez,xu,franco2013dynamics,bellomo2011dynamics,man1,franco2}. The authors would also like to thank C. González-Gutiérrez for providing reference~\cite{gonz}, F. Plastina and T. J. G. Apollaro for providing references~\cite{apol1,apol2}.  SC, AM, and SG gratefully acknowledge discussions with Samyadev Bhattacharya and Chiranjib Mukhopadhyay. 
\end{acknowledgments}

%
%

\bibliography{spin_markov.bib}

\begin{thebibliography}{47}
\expandafter\ifx\csname natexlab\endcsname\relax\def\natexlab#1{#1}\fi
\expandafter\ifx\csname bibnamefont\endcsname\relax
  \def\bibnamefont#1{#1}\fi
\expandafter\ifx\csname bibfnamefont\endcsname\relax
  \def\bibfnamefont#1{#1}\fi
\expandafter\ifx\csname citenamefont\endcsname\relax
  \def\citenamefont#1{#1}\fi
\expandafter\ifx\csname url\endcsname\relax
  \def\url#1{\texttt{#1}}\fi
\expandafter\ifx\csname urlprefix\endcsname\relax\def\urlprefix{URL }\fi
\providecommand{\bibinfo}[2]{#2}
\providecommand{\eprint}[2][]{\url{#2}}

\bibitem[{\citenamefont{Breuer et~al.}(2016)\citenamefont{Breuer, Laine, Piilo,
  and Vacchini}}]{breuer}
\bibinfo{author}{\bibfnamefont{H.-P.} \bibnamefont{Breuer}},
  \bibinfo{author}{\bibfnamefont{E.-M.} \bibnamefont{Laine}},
  \bibinfo{author}{\bibfnamefont{J.}~\bibnamefont{Piilo}}, \bibnamefont{and}
  \bibinfo{author}{\bibfnamefont{B.}~\bibnamefont{Vacchini}},
  \bibinfo{journal}{Reviews of Modern Physics} \textbf{\bibinfo{volume}{88}},
  \bibinfo{pages}{021002} (\bibinfo{year}{2016}).

\bibitem[{\citenamefont{Rivas et~al.}(2014)\citenamefont{Rivas, Huelga, and
  Plenio}}]{rivas}
\bibinfo{author}{\bibfnamefont{A.}~\bibnamefont{Rivas}},
  \bibinfo{author}{\bibfnamefont{S.~F.} \bibnamefont{Huelga}},
  \bibnamefont{and} \bibinfo{author}{\bibfnamefont{M.~B.}
  \bibnamefont{Plenio}}, \bibinfo{journal}{Reports on Progress in Physics}
  \textbf{\bibinfo{volume}{77}}, \bibinfo{pages}{094001}
  (\bibinfo{year}{2014}).

\bibitem[{\citenamefont{Bylicka et~al.}(2014)\citenamefont{Bylicka,
  Chru{\'s}ci{\'n}ski, and Maniscalco}}]{bylicka}
\bibinfo{author}{\bibfnamefont{B.}~\bibnamefont{Bylicka}},
  \bibinfo{author}{\bibfnamefont{D.}~\bibnamefont{Chru{\'s}ci{\'n}ski}},
  \bibnamefont{and}
  \bibinfo{author}{\bibfnamefont{S.}~\bibnamefont{Maniscalco}},
  \bibinfo{journal}{Scientific reports} \textbf{\bibinfo{volume}{4}}
  (\bibinfo{year}{2014}).

\bibitem[{\citenamefont{Man et~al.}(2015{\natexlab{a}})\citenamefont{Man, Xia,
  and Franco}}]{man}
\bibinfo{author}{\bibfnamefont{Z.-X.} \bibnamefont{Man}},
  \bibinfo{author}{\bibfnamefont{Y.-J.} \bibnamefont{Xia}}, \bibnamefont{and}
  \bibinfo{author}{\bibfnamefont{R.~L.} \bibnamefont{Franco}},
  \bibinfo{journal}{Scientific reports} \textbf{\bibinfo{volume}{5}}
  (\bibinfo{year}{2015}{\natexlab{a}}).

\bibitem[{\citenamefont{Franco}(2016)}]{franco1}
\bibinfo{author}{\bibfnamefont{R.~L.} \bibnamefont{Franco}},
  \bibinfo{journal}{Quantum Information Processing}
  \textbf{\bibinfo{volume}{15}}, \bibinfo{pages}{2393} (\bibinfo{year}{2016}).

\bibitem[{\citenamefont{Bellomo et~al.}(2012)\citenamefont{Bellomo, Franco,
  Andersson, Cresser, and Compagno}}]{bellomo}
\bibinfo{author}{\bibfnamefont{B.}~\bibnamefont{Bellomo}},
  \bibinfo{author}{\bibfnamefont{R.~L.} \bibnamefont{Franco}},
  \bibinfo{author}{\bibfnamefont{E.}~\bibnamefont{Andersson}},
  \bibinfo{author}{\bibfnamefont{J.~D.} \bibnamefont{Cresser}},
  \bibnamefont{and} \bibinfo{author}{\bibfnamefont{G.}~\bibnamefont{Compagno}},
  \bibinfo{journal}{Physica Scripta} \textbf{\bibinfo{volume}{2012}},
  \bibinfo{pages}{014004} (\bibinfo{year}{2012}).

\bibitem[{\citenamefont{Gonz{\'a}lez-Guti{\'e}rrez
  et~al.}(2016{\natexlab{a}})\citenamefont{Gonz{\'a}lez-Guti{\'e}rrez,
  Rom{\'a}n-Ancheyta, Espitia, and Lo~Franco}}]{gonzalez}
\bibinfo{author}{\bibfnamefont{C.~A.}
  \bibnamefont{Gonz{\'a}lez-Guti{\'e}rrez}},
  \bibinfo{author}{\bibfnamefont{R.}~\bibnamefont{Rom{\'a}n-Ancheyta}},
  \bibinfo{author}{\bibfnamefont{D.}~\bibnamefont{Espitia}}, \bibnamefont{and}
  \bibinfo{author}{\bibfnamefont{R.}~\bibnamefont{Lo~Franco}},
  \bibinfo{journal}{International Journal of Quantum Information}
  \textbf{\bibinfo{volume}{14}}, \bibinfo{pages}{1650031}
  (\bibinfo{year}{2016}{\natexlab{a}}).

\bibitem[{\citenamefont{Xu et~al.}(2013)\citenamefont{Xu, Sun, Li, Xu, Guo,
  Andersson, Franco, and Compagno}}]{xu}
\bibinfo{author}{\bibfnamefont{J.-S.} \bibnamefont{Xu}},
  \bibinfo{author}{\bibfnamefont{K.}~\bibnamefont{Sun}},
  \bibinfo{author}{\bibfnamefont{C.-F.} \bibnamefont{Li}},
  \bibinfo{author}{\bibfnamefont{X.-Y.} \bibnamefont{Xu}},
  \bibinfo{author}{\bibfnamefont{G.-C.} \bibnamefont{Guo}},
  \bibinfo{author}{\bibfnamefont{E.}~\bibnamefont{Andersson}},
  \bibinfo{author}{\bibfnamefont{R.~L.} \bibnamefont{Franco}},
  \bibnamefont{and} \bibinfo{author}{\bibfnamefont{G.}~\bibnamefont{Compagno}},
  \bibinfo{journal}{Nature communications} \textbf{\bibinfo{volume}{4}}
  (\bibinfo{year}{2013}).

\bibitem[{\citenamefont{Franco et~al.}(2013)\citenamefont{Franco, Bellomo,
  Maniscalco, and Compagno}}]{franco2013dynamics}
\bibinfo{author}{\bibfnamefont{R.~L.} \bibnamefont{Franco}},
  \bibinfo{author}{\bibfnamefont{B.}~\bibnamefont{Bellomo}},
  \bibinfo{author}{\bibfnamefont{S.}~\bibnamefont{Maniscalco}},
  \bibnamefont{and} \bibinfo{author}{\bibfnamefont{G.}~\bibnamefont{Compagno}},
  \bibinfo{journal}{International Journal of Modern Physics B}
  \textbf{\bibinfo{volume}{27}}, \bibinfo{pages}{1345053}
  (\bibinfo{year}{2013}).

\bibitem[{\citenamefont{Bellomo et~al.}(2011)\citenamefont{Bellomo, Compagno,
  Lo~Franco, Ridolfo, and Savasta}}]{bellomo2011dynamics}
\bibinfo{author}{\bibfnamefont{B.}~\bibnamefont{Bellomo}},
  \bibinfo{author}{\bibfnamefont{G.}~\bibnamefont{Compagno}},
  \bibinfo{author}{\bibfnamefont{R.}~\bibnamefont{Lo~Franco}},
  \bibinfo{author}{\bibfnamefont{A.}~\bibnamefont{Ridolfo}}, \bibnamefont{and}
  \bibinfo{author}{\bibfnamefont{S.}~\bibnamefont{Savasta}},
  \bibinfo{journal}{International Journal of Quantum Information}
  \textbf{\bibinfo{volume}{9}}, \bibinfo{pages}{1665} (\bibinfo{year}{2011}).

\bibitem[{\citenamefont{Schmidt et~al.}(1999)\citenamefont{Schmidt, Blaschke,
  R{\"o}pke, Prozorkevich, Smolyansky, and Toneev}}]{schmidt}
\bibinfo{author}{\bibfnamefont{S.}~\bibnamefont{Schmidt}},
  \bibinfo{author}{\bibfnamefont{D.}~\bibnamefont{Blaschke}},
  \bibinfo{author}{\bibfnamefont{G.}~\bibnamefont{R{\"o}pke}},
  \bibinfo{author}{\bibfnamefont{A.}~\bibnamefont{Prozorkevich}},
  \bibinfo{author}{\bibfnamefont{S.}~\bibnamefont{Smolyansky}},
  \bibnamefont{and} \bibinfo{author}{\bibfnamefont{V.}~\bibnamefont{Toneev}},
  \bibinfo{journal}{Physical Review D} \textbf{\bibinfo{volume}{59}},
  \bibinfo{pages}{094005} (\bibinfo{year}{1999}).

\bibitem[{\citenamefont{Bloch et~al.}(2000)\citenamefont{Bloch, Roberts, and
  Schmidt}}]{bloch}
\bibinfo{author}{\bibfnamefont{J.~C.} \bibnamefont{Bloch}},
  \bibinfo{author}{\bibfnamefont{C.~D.} \bibnamefont{Roberts}},
  \bibnamefont{and} \bibinfo{author}{\bibfnamefont{S.}~\bibnamefont{Schmidt}},
  \bibinfo{journal}{Physical Review D} \textbf{\bibinfo{volume}{61}},
  \bibinfo{pages}{117502} (\bibinfo{year}{2000}).

\bibitem[{\citenamefont{Rebentrost and Aspuru-Guzik}(2011)}]{rebentrost}
\bibinfo{author}{\bibfnamefont{P.}~\bibnamefont{Rebentrost}} \bibnamefont{and}
  \bibinfo{author}{\bibfnamefont{A.}~\bibnamefont{Aspuru-Guzik}},
  \emph{\bibinfo{title}{Communication: Exciton--phonon information flow in the
  energy transfer process of photosynthetic complexes}} (\bibinfo{year}{2011}).

\bibitem[{\citenamefont{Liu et~al.}(2011)\citenamefont{Liu, Li, Huang, Li, Guo,
  Laine, Breuer, and Piilo}}]{liu}
\bibinfo{author}{\bibfnamefont{B.-H.} \bibnamefont{Liu}},
  \bibinfo{author}{\bibfnamefont{L.}~\bibnamefont{Li}},
  \bibinfo{author}{\bibfnamefont{Y.-F.} \bibnamefont{Huang}},
  \bibinfo{author}{\bibfnamefont{C.-F.} \bibnamefont{Li}},
  \bibinfo{author}{\bibfnamefont{G.-C.} \bibnamefont{Guo}},
  \bibinfo{author}{\bibfnamefont{E.-M.} \bibnamefont{Laine}},
  \bibinfo{author}{\bibfnamefont{H.-P.} \bibnamefont{Breuer}},
  \bibnamefont{and} \bibinfo{author}{\bibfnamefont{J.}~\bibnamefont{Piilo}},
  \bibinfo{journal}{Nature Physics} \textbf{\bibinfo{volume}{7}},
  \bibinfo{pages}{931} (\bibinfo{year}{2011}).

\bibitem[{\citenamefont{Bernardes et~al.}(2014)\citenamefont{Bernardes,
  Carvalho, Monken, and Santos}}]{bernardes}
\bibinfo{author}{\bibfnamefont{N.}~\bibnamefont{Bernardes}},
  \bibinfo{author}{\bibfnamefont{A.}~\bibnamefont{Carvalho}},
  \bibinfo{author}{\bibfnamefont{C.}~\bibnamefont{Monken}}, \bibnamefont{and}
  \bibinfo{author}{\bibfnamefont{M.~F.} \bibnamefont{Santos}},
  \bibinfo{journal}{Physical Review A} \textbf{\bibinfo{volume}{90}},
  \bibinfo{pages}{032111} (\bibinfo{year}{2014}).

\bibitem[{\citenamefont{Brito and Werlang}(2015)}]{brito}
\bibinfo{author}{\bibfnamefont{F.}~\bibnamefont{Brito}} \bibnamefont{and}
  \bibinfo{author}{\bibfnamefont{T.}~\bibnamefont{Werlang}},
  \bibinfo{journal}{New Journal of Physics} \textbf{\bibinfo{volume}{17}},
  \bibinfo{pages}{072001} (\bibinfo{year}{2015}).

\bibitem[{\citenamefont{Garrido et~al.}(2016)\citenamefont{Garrido, Gorin, and
  Pineda}}]{garrido}
\bibinfo{author}{\bibfnamefont{N.}~\bibnamefont{Garrido}},
  \bibinfo{author}{\bibfnamefont{T.}~\bibnamefont{Gorin}}, \bibnamefont{and}
  \bibinfo{author}{\bibfnamefont{C.}~\bibnamefont{Pineda}},
  \bibinfo{journal}{Physical Review A} \textbf{\bibinfo{volume}{93}},
  \bibinfo{pages}{012113} (\bibinfo{year}{2016}).

\bibitem[{\citenamefont{Man et~al.}(2015{\natexlab{b}})\citenamefont{Man, Xia,
  and Franco}}]{man1}
\bibinfo{author}{\bibfnamefont{Z.-X.} \bibnamefont{Man}},
  \bibinfo{author}{\bibfnamefont{Y.-J.} \bibnamefont{Xia}}, \bibnamefont{and}
  \bibinfo{author}{\bibfnamefont{R.~L.} \bibnamefont{Franco}},
  \bibinfo{journal}{Physical Review A} \textbf{\bibinfo{volume}{92}},
  \bibinfo{pages}{012315} (\bibinfo{year}{2015}{\natexlab{b}}).

\bibitem[{\citenamefont{Franco}(2015)}]{franco2}
\bibinfo{author}{\bibfnamefont{R.~L.} \bibnamefont{Franco}},
  \bibinfo{journal}{New Journal of Physics} \textbf{\bibinfo{volume}{17}},
  \bibinfo{pages}{081004} (\bibinfo{year}{2015}).

\bibitem[{\citenamefont{Prokof'ev and Stamp}(2000)}]{spinbath}
\bibinfo{author}{\bibfnamefont{N.}~\bibnamefont{Prokof'ev}} \bibnamefont{and}
  \bibinfo{author}{\bibfnamefont{P.}~\bibnamefont{Stamp}},
  \bibinfo{journal}{Reports on Progress in Physics}
  \textbf{\bibinfo{volume}{63}}, \bibinfo{pages}{669} (\bibinfo{year}{2000}).

\bibitem[{\citenamefont{Hutton and Bose}(2004)}]{hutton2004mediated}
\bibinfo{author}{\bibfnamefont{A.}~\bibnamefont{Hutton}} \bibnamefont{and}
  \bibinfo{author}{\bibfnamefont{S.}~\bibnamefont{Bose}},
  \bibinfo{journal}{Physical Review A} \textbf{\bibinfo{volume}{69}},
  \bibinfo{pages}{042312} (\bibinfo{year}{2004}).

\bibitem[{\citenamefont{Breuer et~al.}(2004)\citenamefont{Breuer, Burgarth, and
  Petruccione}}]{breuer2004non}
\bibinfo{author}{\bibfnamefont{H.-P.} \bibnamefont{Breuer}},
  \bibinfo{author}{\bibfnamefont{D.}~\bibnamefont{Burgarth}}, \bibnamefont{and}
  \bibinfo{author}{\bibfnamefont{F.}~\bibnamefont{Petruccione}},
  \bibinfo{journal}{Physical Review B} \textbf{\bibinfo{volume}{70}},
  \bibinfo{pages}{045323} (\bibinfo{year}{2004}).

\bibitem[{\citenamefont{Bhattacharya et~al.}(2017)\citenamefont{Bhattacharya,
  Misra, Mukhopadhyay, and Pati}}]{samyadev}
\bibinfo{author}{\bibfnamefont{S.}~\bibnamefont{Bhattacharya}},
  \bibinfo{author}{\bibfnamefont{A.}~\bibnamefont{Misra}},
  \bibinfo{author}{\bibfnamefont{C.}~\bibnamefont{Mukhopadhyay}},
  \bibnamefont{and} \bibinfo{author}{\bibfnamefont{A.~K.} \bibnamefont{Pati}},
  \bibinfo{journal}{Physical Review A} \textbf{\bibinfo{volume}{95}},
  \bibinfo{pages}{012122} (\bibinfo{year}{2017}).

\bibitem[{\citenamefont{Jing and Wu}(2018)}]{jing}
\bibinfo{author}{\bibfnamefont{J.}~\bibnamefont{Jing}} \bibnamefont{and}
  \bibinfo{author}{\bibfnamefont{L.-A.} \bibnamefont{Wu}},
  \bibinfo{journal}{Scientific reports} \textbf{\bibinfo{volume}{8}},
  \bibinfo{pages}{1471} (\bibinfo{year}{2018}).

\bibitem[{\citenamefont{Rivas et~al.}(2010)\citenamefont{Rivas, Huelga, and
  Plenio}}]{rivas2}
\bibinfo{author}{\bibfnamefont{{\'A}.}~\bibnamefont{Rivas}},
  \bibinfo{author}{\bibfnamefont{S.~F.} \bibnamefont{Huelga}},
  \bibnamefont{and} \bibinfo{author}{\bibfnamefont{M.~B.}
  \bibnamefont{Plenio}}, \bibinfo{journal}{Physical review letters}
  \textbf{\bibinfo{volume}{105}}, \bibinfo{pages}{050403}
  (\bibinfo{year}{2010}).

\bibitem[{\citenamefont{Apollaro et~al.}(2011)\citenamefont{Apollaro,
  Di~Franco, Plastina, and Paternostro}}]{apol1}
\bibinfo{author}{\bibfnamefont{T.~J.} \bibnamefont{Apollaro}},
  \bibinfo{author}{\bibfnamefont{C.}~\bibnamefont{Di~Franco}},
  \bibinfo{author}{\bibfnamefont{F.}~\bibnamefont{Plastina}}, \bibnamefont{and}
  \bibinfo{author}{\bibfnamefont{M.}~\bibnamefont{Paternostro}},
  \bibinfo{journal}{Physical Review A} \textbf{\bibinfo{volume}{83}},
  \bibinfo{pages}{032103} (\bibinfo{year}{2011}).

\bibitem[{\citenamefont{Lorenzo et~al.}(2013)\citenamefont{Lorenzo, Plastina,
  and Paternostro}}]{apol2}
\bibinfo{author}{\bibfnamefont{S.}~\bibnamefont{Lorenzo}},
  \bibinfo{author}{\bibfnamefont{F.}~\bibnamefont{Plastina}}, \bibnamefont{and}
  \bibinfo{author}{\bibfnamefont{M.}~\bibnamefont{Paternostro}},
  \bibinfo{journal}{Physical Review A} \textbf{\bibinfo{volume}{87}},
  \bibinfo{pages}{022317} (\bibinfo{year}{2013}).

\bibitem[{\citenamefont{Wang et~al.}(2013)\citenamefont{Wang, Guo, and
  Zhou}}]{wang2013non}
\bibinfo{author}{\bibfnamefont{Z.}~\bibnamefont{Wang}},
  \bibinfo{author}{\bibfnamefont{Y.}~\bibnamefont{Guo}}, \bibnamefont{and}
  \bibinfo{author}{\bibfnamefont{D.}~\bibnamefont{Zhou}}, \bibinfo{journal}{The
  European Physical Journal D} \textbf{\bibinfo{volume}{67}},
  \bibinfo{pages}{218} (\bibinfo{year}{2013}).

\bibitem[{\citenamefont{Breuer et~al.}(2009)\citenamefont{Breuer, Laine, and
  Piilo}}]{breuer1}
\bibinfo{author}{\bibfnamefont{H.-P.} \bibnamefont{Breuer}},
  \bibinfo{author}{\bibfnamefont{E.-M.} \bibnamefont{Laine}}, \bibnamefont{and}
  \bibinfo{author}{\bibfnamefont{J.}~\bibnamefont{Piilo}},
  \bibinfo{journal}{Physical review letters} \textbf{\bibinfo{volume}{103}},
  \bibinfo{pages}{210401} (\bibinfo{year}{2009}).

\bibitem[{\citenamefont{Wi{\ss}mann et~al.}(2015)\citenamefont{Wi{\ss}mann,
  Breuer, and Vacchini}}]{wise}
\bibinfo{author}{\bibfnamefont{S.}~\bibnamefont{Wi{\ss}mann}},
  \bibinfo{author}{\bibfnamefont{H.-P.} \bibnamefont{Breuer}},
  \bibnamefont{and} \bibinfo{author}{\bibfnamefont{B.}~\bibnamefont{Vacchini}},
  \bibinfo{journal}{Physical Review A} \textbf{\bibinfo{volume}{92}},
  \bibinfo{pages}{042108} (\bibinfo{year}{2015}).

\bibitem[{\citenamefont{Luo et~al.}(2012)\citenamefont{Luo, Fu, and
  Song}}]{luo}
\bibinfo{author}{\bibfnamefont{S.}~\bibnamefont{Luo}},
  \bibinfo{author}{\bibfnamefont{S.}~\bibnamefont{Fu}}, \bibnamefont{and}
  \bibinfo{author}{\bibfnamefont{H.}~\bibnamefont{Song}},
  \bibinfo{journal}{Physical Review A} \textbf{\bibinfo{volume}{86}},
  \bibinfo{pages}{044101} (\bibinfo{year}{2012}).

\bibitem[{\citenamefont{Bylicka et~al.}(2017)\citenamefont{Bylicka, Johansson,
  and Ac{\'\i}n}}]{bylicka2017constructive}
\bibinfo{author}{\bibfnamefont{B.}~\bibnamefont{Bylicka}},
  \bibinfo{author}{\bibfnamefont{M.}~\bibnamefont{Johansson}},
  \bibnamefont{and}
  \bibinfo{author}{\bibfnamefont{A.}~\bibnamefont{Ac{\'\i}n}},
  \bibinfo{journal}{Physical review letters} \textbf{\bibinfo{volume}{118}},
  \bibinfo{pages}{120501} (\bibinfo{year}{2017}).

\bibitem[{\citenamefont{Chru{\'s}ci{\'n}ski and
  Rivas}(2017)}]{chruscinski2017universal}
\bibinfo{author}{\bibfnamefont{D.}~\bibnamefont{Chru{\'s}ci{\'n}ski}}
  \bibnamefont{and}
  \bibinfo{author}{\bibfnamefont{{\'A}.}~\bibnamefont{Rivas}},
  \bibinfo{journal}{arXiv preprint arXiv:1710.06771}  (\bibinfo{year}{2017}).

\bibitem[{\citenamefont{Chakraborty}(2018)}]{chakraborty}
\bibinfo{author}{\bibfnamefont{S.}~\bibnamefont{Chakraborty}},
  \bibinfo{journal}{Physical Review A} \textbf{\bibinfo{volume}{97}},
  \bibinfo{pages}{032130} (\bibinfo{year}{2018}).

\bibitem[{\citenamefont{Chru{\'s}ci{\'n}ski and Kossakowski}(2012)}]{chru}
\bibinfo{author}{\bibfnamefont{D.}~\bibnamefont{Chru{\'s}ci{\'n}ski}}
  \bibnamefont{and}
  \bibinfo{author}{\bibfnamefont{A.}~\bibnamefont{Kossakowski}},
  \bibinfo{journal}{Journal of Physics B: Atomic, Molecular and Optical
  Physics} \textbf{\bibinfo{volume}{45}}, \bibinfo{pages}{154002}
  (\bibinfo{year}{2012}).

\bibitem[{\citenamefont{Breuer and Petruccione}(2002)}]{breuer2002theory}
\bibinfo{author}{\bibfnamefont{H.-P.} \bibnamefont{Breuer}} \bibnamefont{and}
  \bibinfo{author}{\bibfnamefont{F.}~\bibnamefont{Petruccione}},
  \emph{\bibinfo{title}{The theory of open quantum systems}}
  (\bibinfo{publisher}{Oxford University Press on Demand},
  \bibinfo{year}{2002}).

\bibitem[{\citenamefont{Peres}(1996)}]{peres}
\bibinfo{author}{\bibfnamefont{A.}~\bibnamefont{Peres}},
  \bibinfo{journal}{Physical Review Letters} \textbf{\bibinfo{volume}{77}},
  \bibinfo{pages}{1413} (\bibinfo{year}{1996}).

\bibitem[{\citenamefont{Horodecki et~al.}(1996)\citenamefont{Horodecki,
  Horodecki, and Horodecki}}]{horod}
\bibinfo{author}{\bibfnamefont{M.}~\bibnamefont{Horodecki}},
  \bibinfo{author}{\bibfnamefont{P.}~\bibnamefont{Horodecki}},
  \bibnamefont{and}
  \bibinfo{author}{\bibfnamefont{R.}~\bibnamefont{Horodecki}},
  \bibinfo{journal}{Physics Letters A} \textbf{\bibinfo{volume}{223}},
  \bibinfo{pages}{1 } (\bibinfo{year}{1996}).

\bibitem[{\citenamefont{Sanpera et~al.}(1998)\citenamefont{Sanpera, Tarrach,
  and Vidal}}]{sanpera1998local}
\bibinfo{author}{\bibfnamefont{A.}~\bibnamefont{Sanpera}},
  \bibinfo{author}{\bibfnamefont{R.}~\bibnamefont{Tarrach}}, \bibnamefont{and}
  \bibinfo{author}{\bibfnamefont{G.}~\bibnamefont{Vidal}},
  \bibinfo{journal}{Physical Review A} \textbf{\bibinfo{volume}{58}},
  \bibinfo{pages}{826} (\bibinfo{year}{1998}).

\bibitem[{\citenamefont{Rana and Parashar}(2012)}]{rana2012entanglement}
\bibinfo{author}{\bibfnamefont{S.}~\bibnamefont{Rana}} \bibnamefont{and}
  \bibinfo{author}{\bibfnamefont{P.}~\bibnamefont{Parashar}},
  \bibinfo{journal}{Physical Review A} \textbf{\bibinfo{volume}{86}},
  \bibinfo{pages}{030302} (\bibinfo{year}{2012}).

\bibitem[{\citenamefont{Vidal and Werner}(2002)}]{vidal2002computable}
\bibinfo{author}{\bibfnamefont{G.}~\bibnamefont{Vidal}} \bibnamefont{and}
  \bibinfo{author}{\bibfnamefont{R.~F.} \bibnamefont{Werner}},
  \bibinfo{journal}{Physical Review A} \textbf{\bibinfo{volume}{65}},
  \bibinfo{pages}{032314} (\bibinfo{year}{2002}).

\bibitem[{\citenamefont{Sadiek et~al.}(2010)\citenamefont{Sadiek, Alkurtass,
  and Aldossary}}]{sadiek2010entanglement}
\bibinfo{author}{\bibfnamefont{G.}~\bibnamefont{Sadiek}},
  \bibinfo{author}{\bibfnamefont{B.}~\bibnamefont{Alkurtass}},
  \bibnamefont{and}
  \bibinfo{author}{\bibfnamefont{O.}~\bibnamefont{Aldossary}},
  \bibinfo{journal}{Physical Review A} \textbf{\bibinfo{volume}{82}},
  \bibinfo{pages}{052337} (\bibinfo{year}{2010}).

\bibitem[{\citenamefont{Bortz and Stolze}(2007)}]{bortz2007spin}
\bibinfo{author}{\bibfnamefont{M.}~\bibnamefont{Bortz}} \bibnamefont{and}
  \bibinfo{author}{\bibfnamefont{J.}~\bibnamefont{Stolze}},
  \bibinfo{journal}{Journal of Statistical Mechanics: Theory and Experiment}
  \textbf{\bibinfo{volume}{2007}}, \bibinfo{pages}{P06018}
  (\bibinfo{year}{2007}).

\bibitem[{\citenamefont{Hoffman and Kunze}()}]{hoffmanlinear}
\bibinfo{author}{\bibfnamefont{K.}~\bibnamefont{Hoffman}} \bibnamefont{and}
  \bibinfo{author}{\bibfnamefont{R.}~\bibnamefont{Kunze}},
  \bibinfo{journal}{Englewood Cliffs, New Jersey}  (????).

\bibitem[{\citenamefont{Dive et~al.}(2015)\citenamefont{Dive, Mintert, and
  Burgarth}}]{dive2015quantum}
\bibinfo{author}{\bibfnamefont{B.}~\bibnamefont{Dive}},
  \bibinfo{author}{\bibfnamefont{F.}~\bibnamefont{Mintert}}, \bibnamefont{and}
  \bibinfo{author}{\bibfnamefont{D.}~\bibnamefont{Burgarth}},
  \bibinfo{journal}{Physical Review A} \textbf{\bibinfo{volume}{92}},
  \bibinfo{pages}{032111} (\bibinfo{year}{2015}).

\bibitem[{\citenamefont{Pang et~al.}(2017)\citenamefont{Pang, Brun, and
  Jordan}}]{pang2017abrupt}
\bibinfo{author}{\bibfnamefont{S.}~\bibnamefont{Pang}},
  \bibinfo{author}{\bibfnamefont{T.~A.} \bibnamefont{Brun}}, \bibnamefont{and}
  \bibinfo{author}{\bibfnamefont{A.~N.} \bibnamefont{Jordan}},
  \bibinfo{journal}{arXiv preprint arXiv:1712.10109}  (\bibinfo{year}{2017}).

\bibitem[{\citenamefont{Gonz{\'a}lez-Guti{\'e}rrez
  et~al.}(2016{\natexlab{b}})\citenamefont{Gonz{\'a}lez-Guti{\'e}rrez,
  Villase{\~n}or, Pineda, and Seligman}}]{gonz}
\bibinfo{author}{\bibfnamefont{C.}~\bibnamefont{Gonz{\'a}lez-Guti{\'e}rrez}},
  \bibinfo{author}{\bibfnamefont{E.}~\bibnamefont{Villase{\~n}or}},
  \bibinfo{author}{\bibfnamefont{C.}~\bibnamefont{Pineda}}, \bibnamefont{and}
  \bibinfo{author}{\bibfnamefont{T.}~\bibnamefont{Seligman}},
  \bibinfo{journal}{Physica Scripta} \textbf{\bibinfo{volume}{91}},
  \bibinfo{pages}{083001} (\bibinfo{year}{2016}{\natexlab{b}}).

\end{thebibliography}

\end{document}